\documentclass{emulateapj}
\usepackage{graphicx}    
\usepackage{subfigure}
\usepackage{color}
\newcommand{\Msun}{M_{\odot}}
\newcommand{\Mvir}{M_{\rm DM}}
\newcommand{\Mgas}{M_{\rm gas}}
\newcommand{\Mstar}{M_{\rm star}}

\newcommand{\rDM}{(m/M)_{\rm DM}}
\newcommand{\rstar}{(m/M)_{\rm star}}
\newcommand{\rgal}{(m/M)_{\rm gal}}
\submitted{}

\begin{document}
\title{Gas-Rich Mergers in LCDM: Disk Survivability and the Baryonic Assembly of Galaxies}
\author{Kyle R. Stewart\altaffilmark{1}, James S. Bullock\altaffilmark{1},
Risa H. Wechsler\altaffilmark{2}, and Ariyeh H. Maller\altaffilmark{3}}

\altaffiltext{1}{Center for Cosmology, Department of Physics and Astronomy, The University of California at Irvine, Irvine, CA, 92697, USA}
\altaffiltext{2}{Kavli Institute for Particle Astrophysics \& Cosmology, Physics Department, and Stanford Linear Accelerator Center, Stanford University, Stanford, CA 94305, USA}
\altaffiltext{3}{Department of Physics, New York City College of Technology, 300 Jay St., Brooklyn, NY 11201, USA}

\begin{abstract} {
We use $N$-body simulations and observationally-normalized relations
between dark matter halo mass, stellar mass, and cold gas mass to derive robust
expectations about the baryonic content of major mergers out to
redshift $z\sim2$.
First, we find that the majority of major mergers ($m/M > 0.3 $) experienced by
Milky Way size dark matter halos  should have been gas-rich, and that gas-rich mergers are increasingly
common at high redshift.
Though the frequency of major mergers
into galaxy  halos in our simulations greatly exceeds the observed
early-type galaxy fraction, the frequency of {\em gas-poor} major
 mergers is consistent with the observed
fraction of bulge-dominated galaxies
across the  halo mass range $\Mvir\sim10^{11}-10^{13}\Msun$.
These results lend support to the conjecture that
mergers with high baryonic gas fractions play an important role in building and/or preserving disk
galaxies in the universe.
Secondly, we find that there is a transition mass
below which a galaxy's past major mergers
were primarily gas-rich and above which they were gas poor.
The associated stellar mass scale corresponds closely to that marking the observed
bimodal division between blue,
star-forming, disk-dominated systems and red, bulge-dominated systems
with old populations.
Finally, we find that the overall fraction of a galaxy's cold baryons deposited directly via major mergers
is significant.  Approximately $\sim20-30\%$ of the cold baryonic material in $\Mstar \sim 10^{10.5} M_{\odot}$
($\Mvir \sim 10^{12} \Msun$) galaxies is accreted as cold gas or stars via major mergers since $z=2$, with
most of this accretion in the form of cold gas.
For more massive galaxies with $\Mstar \sim 10^{11} M_\odot$ ($\Mvir \sim 10^{13} \Msun$) the
fraction of baryons amassed in mergers since $z=2$ is even higher, $\sim40\%$, but most of these accreted baryons are
delivered directly in the form of stars.
This baryonic mass deposition is almost unavoidable, and provides a limit on the
fraction of a galaxy's cold baryons that can originate in cold flows
or from hot halo cooling.
}

\end{abstract}

\section{Introduction}
\label{Introduction}

In the cold dark matter (CDM) model of structure formation, major galaxy mergers
are believed to play an important role in determining a galaxy's morphology
\citep[e.g.][]{ToomreToomre, BarnesHernquist96,Robertson06a,Robertson06b,Burkert08}, as well as
triggering star formation and AGN activity \citep[e.g.][]{MihosHernquist96,Heckman86, Springel05b,cox07},
while minor mergers may help explain
the origin of thick disks and extended diffuse light components around galaxies
\citep{BarnesHernquist96,k07,Purcell07,Younger07,Purcell08,VillalobosHelmi08,Kazantzidis08}.
More than simply triggering star formation in existing gas and altering existing galaxy
morphologies, mergers  deliver new stars and additional fuel for star formation, and thereby contribute to
baryonic acquisition of galaxies over their histories.
That mergers contribute significantly to many aspects of galaxy formation is now fairly well accepted, however
there are lingering concerns that mergers are too common in CDM
to explain the prominence of thin disk-dominated galaxies in the local universe
\citep[e.g.][and references therein]{toth_ostriker92,walker_etal96,Stewart08,Purcell08b,BSP08}.
Here we explore the baryonic content of these predicted mergers and the potential ramifications
of gas-rich and gas-poor mergers on galactic morphological evolution.

The baryonic delivery of material into galaxies via major mergers touches on a broader
question in galaxy formation: how do galaxies get their baryons?
In recent years, studies motivated by hydrodynamic simulations have placed a growing emphasis on
the importance of smooth gas accretion via ``cold flows.''  These cold flows
constitute streams of cold gas flowing along
filamentary structures (particularly at high redshift)
with sufficiently high densities to penetrate into a halo's central region without
heating the gas to the virial temperature
\citep[e.g.][]{BirnboimDekel03,Keres05,DekelBirnboim06,Keres08,Dekel08,Brooks08,Agertz09}.
These simulations demonstrate a characteristic halo mass scale ($\sim 10^{12} \Msun$) below which
cold streams are the dominant mode of gas accretion, and above which gas cooling directly from shock-heated (hot mode) material
dominates.

Though there has yet to be any observational evidence that cold flows actually occur in nature, the possibility is well motivated by
theory and is suggestive of a number of interesting scenarios for galaxy assembly. One particularly
interesting idea is that flows of cold gas are vital to the formation of disk galaxies at high redshift $z\gtrsim1$ \citep{Dekel08}.
Still, even if disks were built at high redshift via streams of cold gas, we can return to the issue of disk {\em survival}
raised above.  How do observed populations of disk galaxies at high redshift
\citep[e.g.][at $z\sim1.6$]{Wright08} survive subsequent mergers and
remain disk-dominated by $z=0$?

In a previous paper \citep{Stewart08} we
studied the merger histories of Milky Way-size dark matter
halos within a cosmological $N$-body simulation
and found that approximately $70 \%$ should have
accreted an object with more than twice the mass of the Milky Way disk ($m > 10^{11}
\Msun$) in the last 10 Gyr. In order to achieve the $\sim 70 \%$
disk-dominated fraction that has been observed in Milky Way-sized halos
\citep{Weinmann06,Park07,Choi07,Weinmann09}, mergers involving $>1/3$ mass-ratio events
must \emph{not always} destroy disks.  Adding to the associated concern,
Purcell, Kazantzidis, and Bullock (2008b) performed focused numerical experiments
to study the impact of $m = 10^{11} \Msun$ encounters onto fully-formed Milky-Way type thin {\em stellar} disks
and concluded that {\em thin} ($\sim 400$ pc) dissipationless stellar disks do not survive these (presumably common)
encounters.

\begin{figure*}[th!]
  \includegraphics[width=0.325\textwidth]{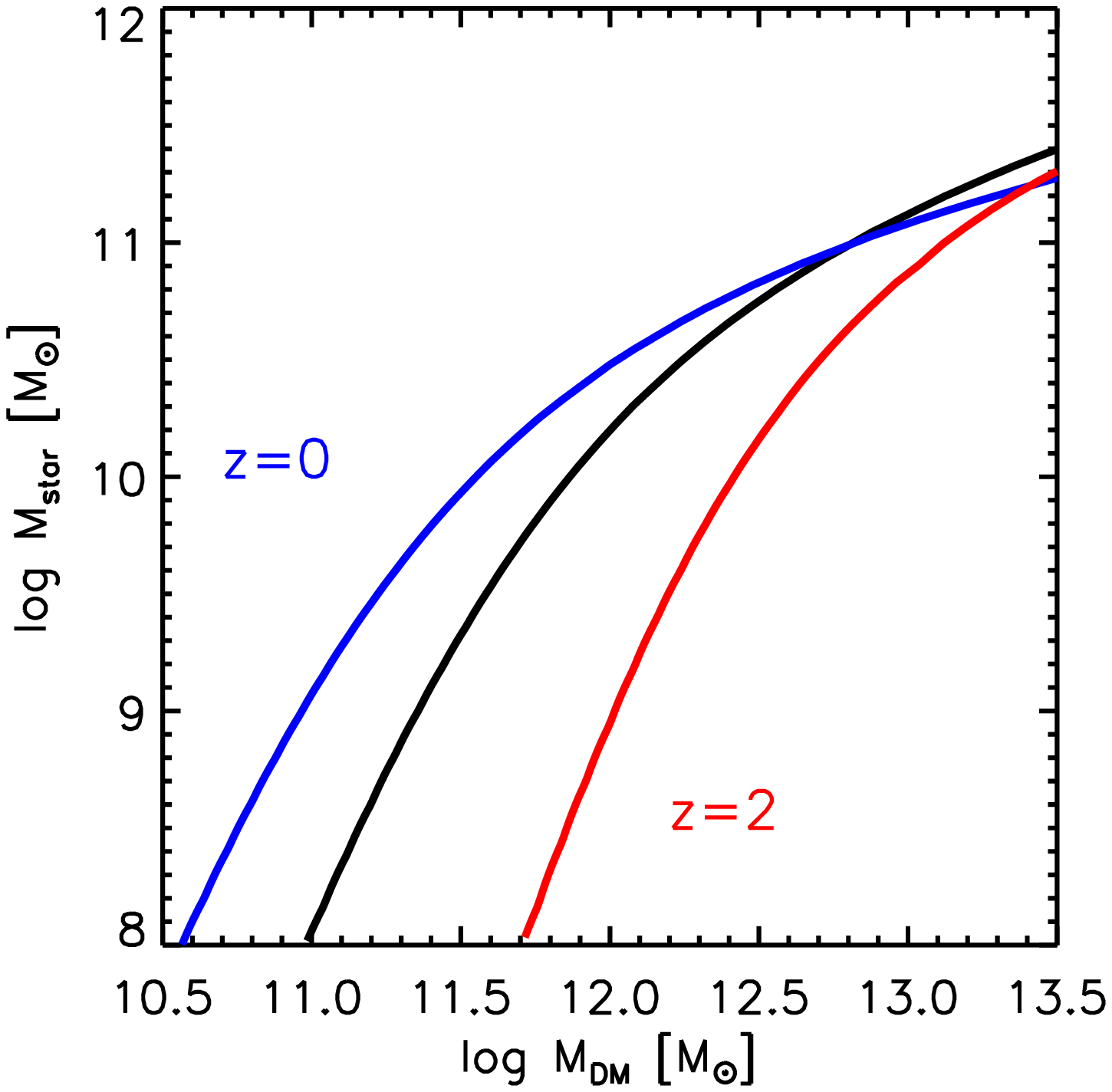}
  \includegraphics[width=0.64\textwidth]{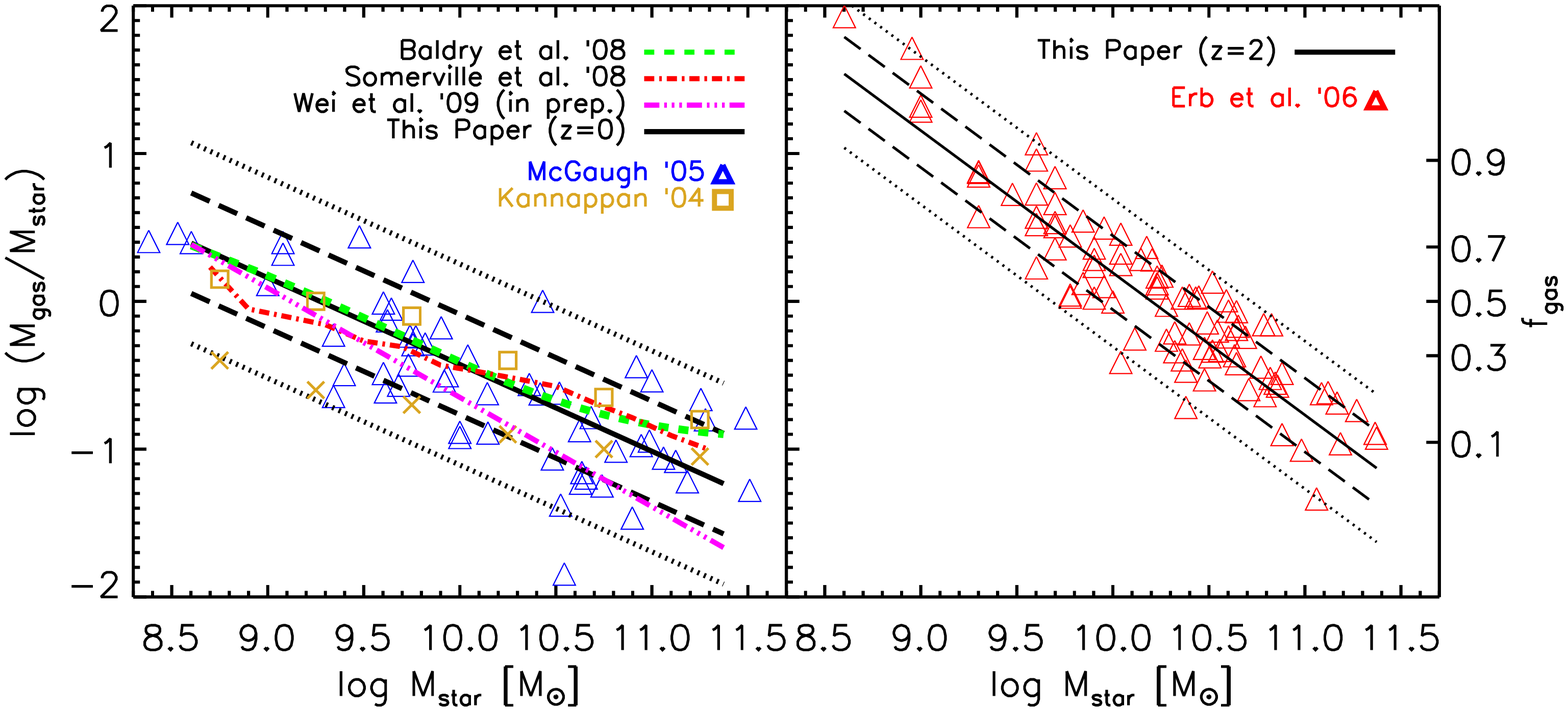}
    \caption{Two step method for assigning baryons to dark matter halos.
\emph{Left:} Stellar mass $\Mstar$ versus dark matter halo mass $\Mvir$, for
$z=0,1,2$, based on abundance matching \citep{ConroyWechsler08}.
\emph{Middle(Right):} Power law fits to $\Mgas/\Mstar$
as a function of stellar mass at $z=0$($2$), with the corresponding
gas fraction $f_{\rm gas}=\Mgas/(\Mgas+\Mstar)$ shown on the right axis.
The symbols represent observational estimates
from \citealp{McGaugh05} (entire sample: blue triangles),
\citealp{Kannappan04} (average of blue galaxy sample, gold squares; average of
red galaxy sample, gold X's),
\citealp{Baldry08} (average: green short-dashed line), \citealp{Somerville08} (average: red dot-dashed line),
\citealp{Wei09} (best fit to total sample: magenta dot-dot-dot-dashed line) and  \citealp{Erb06} (entire sample: red triangles).
The solid (black) line in each panel is the best-fit relation (Equation \ref{eq:gasmass}),
while the long-dashed and dotted (black) lines represent the
$1\sigma$ and $2\sigma$ scatter, respectively.}
\label{McGaughErbfits}
\end{figure*}

One possible solution appeals to the role of cold gas in mergers.  Focused merger simulations in the
past few years have begun to suggest that sufficiently gas-rich mergers
may help build angular momentum in the
central galaxy, while feedback physics prevents the gas from forming stars too quickly,
resulting in a disk-dominated merger remnant \citep{Barnes02,SpringelHernquist05,Robertson06a,Brook07a,Hopkins08g,RobertsonBullock08}.
Encouragingly, cosmological simulations
that have been successful in reproducing disk galaxies have also shown that gas-rich mergers
have played an important role in the disk's creation
\citep[e.g.][]{Brook04,Governato07,Governato08}, and there
have been recent observations of late-type galaxies that may be in the process of
reforming after a recent gas-rich major merger \citep[e.g.][]{Hammer09a}.
\cite{RobertsonBullock08} also showed that the disk-like
merger remnants from gas-rich mergers are similar
to the kinematically hot disks that have been observed at
$z\sim2$ \citep{Schreiber06,Genzel06,Shapiro08}.

Our goal in this work is to provide an empirically-motivated accounting for the expected gas and stellar content of mergers by
relying on robustly determined dark matter halo merger rates.  We aim first to determine whether
gas-rich mergers are common
enough to significantly alleviate the disk formation problem (a necessary but not sufficient condition in evaluating this scenario).
We also aim to investigate the overall importance that mergers
play in the acquisition of a galaxy's cold baryons.

While the merger histories of the dark matter halos are
predicted accurately in dissipationless $N$-body simulations,
the baryonic content of these mergers are much more difficult to predict from first principles.  Indeed,
an accurate {\em ab initio} accounting of the baryonic content of dark matter halos would require an
overarching theory that solved all of the major problems in galaxy formation, including star formation, feedback, and the complicated interplay between
mergers and galaxy assembly.
In this work we chose to avoid the issues of galaxy formation physics entirely.
Instead we adopt a semi-empirical approach that forces our model to match observations at
various redshifts.
First, we adopt the technique of
monotonic abundance matching to assign stellar masses to dark matter halos \citep[specifically following][]{ConroyWechsler08}.
Second, we use observational relations between stellar mass and gas mass \citep[e.g.][]{McGaugh05, Erb06}
to assign gas masses to our halos.  We then combine these relations with the
$N$-body halo merger histories described in \cite{Stewart08} and \cite{Stewart09a}
in order to determine the baryonic properties of mergers back to redshift $z \sim 2$.
In \S2 we discuss the details of our method.
We present our primary results and discuss the implications for
disk survival and the baryonic assembly of galaxies via mergers in \S3.
We summarize our main conclusions in \S4.

\section{Method}
\subsection{The Simulation}

Our simulation contains $512^{3}$ particles, each  with mass $m_p =
3.16 \times 10^8 \Msun$, evolved within a comoving cubic volume of $80
h^{-1}$ Mpc on  a side using    the Adaptive  Refinement  Tree (ART)
$N$-body code \citep{Kravtsov97,Kravtsov04a}.
We assume LCDM cosmological parameters:
$\Omega_{M}=1-\Omega_{\Lambda}=0.3$, $h=0.7$,  and   $\sigma_{8}=0.9$.
The simulation root   computational  grid consists of  $512^3$  cells,
which are  adaptively refined to  a maximum of eight levels, resulting
in a  peak spatial resolution of  $1.2 h^{-1}$ kpc (comoving).
We give only a brief overview of the simulation here, as it has been
reviewed more extensively in previous works.
We refer the reader to \cite{Stewart08}, and reference therein, for a
more complete discussion.

Field dark matter halos and subhalos are identified using a variant of
the bound density maxima algorithm \citep{Klypin99}.  A \emph{subhalo}
is defined as a dark matter halo whose center is positioned within the
virial radius of a more massive halo, whereas a
\emph{field halo} does not.  The virial radius is defined as the radius
of a collapsed self gravitating dark matter halo within which the
average density is $\Delta_{\rm vir}$ times the mean density of the
universe.  Under comparison to constructed mass functions, we have determined that
our halo catalogs are complete to a minimum mass of $10^{10} \Msun$, and our
sample includes $\sim17,000$ field halos at $z=0$ in the mass range
$M=10^{11.2-13.2}\Msun$.  We use the same merger trees described in \cite{Stewart08},
constructed using the techniques described in \cite{Wechsler02, Wechsler06}.

We present our results primarily in terms of the
dark matter mass ratio between the two halos that are undergoing a merger, $\rDM$, where
we always define $m_{\rm DM}$ as the mass of the smaller dark matter halo
(which we will sometimes refer to as the \emph{satellite} halo) just \emph{prior} to
entering the virial radius of the larger one, and $\Mvir$ is the
mass of the larger dark matter halo (also referred to as the \emph{host} halo) at this infall epoch.
Thus, $\Mvir$ does not incorporate the mass
$m_{\rm DM},$ and $\rDM$ has a maximum value of $1.0$.
However, we also present results in terms of
the \emph{stellar} mass ratio of the central galaxies within merging halos, or the mass ratio
between the \emph{total} baryonic mass of these central galaxies
(stellar mass plus gas mass).  We refer to the dark matter, stellar, and galaxy (baryonic) mass ratios
as $\rDM, \rstar, \rgal,$ respectively.
Independent of the mass ratio definitions above, we always refer to a merger ratio of $m/M > 0.3$ as a \emph{major merger}.
For the sake of comparison to our past work, we emphasize
that in \cite{Stewart08}, we considered two definitions of merger ratio.
The first (written as $m/M_0$ in that paper) referred to the ratio of the satellite mass at infall $m$ to the
final dark matter halo mass $M_0$ at $z=0$.  This is \emph{not} the ratio we are using here.  The mass ratio
definition we adopt here is more standard and refers always to the mass ratio at the redshift $z$ of accretion ($m/M_z$ in the notation of
Stewart et al. 2008).

In order to provide robust results, we define a merger to occur once the smaller halo crosses within the
virial radius of the larger halo and becomes a subhalo, as the subsequent orbital evolution of each
subhalo will depend on the baryonic distribution within both halos.
We emphasize that for the major mergers we consider in this paper, the dynamical friction decay timescales
are expected to be short (comparable to the halo dynamical timescale) for typical orbital parameters \citep{BoylanKolchin08}.
Since only $\sim5\%$ of $10^{12}\Msun$ halos have experienced a major merger in the past halo
dynamical time at $z=0$ (see Figure \ref{fractau}), most of these major mergers
into the virial radius have had adequate time to impact the central galaxy and do not survive as
distinct substructure by $z=0$.  For a more in-depth comparison between merger rates into
the virial radius and the rate at which accreted satellites are ``destroyed'' in this simulation
(i.e.~once they lose $90\%$ of their infall mass) we refer the reader to \cite{Stewart09a}.

\subsection{Assigning Stars}

One particularly simple, yet surprisingly successful
approach for assigning galaxies to dark matter halos
is to assume a monotonic mapping between dark matter halo mass $\Mvir$ and
galaxy luminosity $L$ \citep{Kravtsov04a, Tasitsiomi04, ValeOstriker04,
Conroy06, Berrier06, Purcell07, Marin08}.
Using this technique, provided we know $n_g(>L)$ (the cumulative number
density of galaxies brighter than $L$)
we may determine the associated dark matter halo population
by finding the halo mass above which the number density of halos
(including subhalos) matches that of the galaxy population, $n_h(>\Mvir) = n_g(>L)$.

We use a similar approach, and instead assume
a monotonic relationship between halo mass and stellar mass $\Mstar$.
Specifically, we adopt the relation found by
\cite{ConroyWechsler08} (hereafter CW08; interpolated from their data as shown in their Figure 2).
Figure \ref{McGaughErbfits} (left panel)
shows the resulting relation
between stellar mass and dark matter halo mass for $z=0,1,2$ (upper blue, middle black,
lower red curves respectively) where  $\Mstar$ is the stellar mass of the central galaxy
residing within a dark matter halo of mass $\Mvir$.
We ignore scatter in this relationship for the results that follow, but we find that including a Gaussian scatter of $0.1$
dex has no substantive effect our results.

 Of course, a simple relation of this kind
cannot be correct in detail, however, in an average sense, it provides a good characterization
of the relationship between halo mass and galaxy stellar mass that must hold in order for
LCDM to reproduce the observed universe.  Moreover, by adopting it we insure that our model
self-consistently reproduces the observed stellar mass function of galaxies out to $z\sim2$.
We cannot use this method to explore merger rates as a function of stellar mass beyond $z\sim 2$
because the stellar mass function is poorly constrained at higher redshifts.
We refer the reader to \cite{Marchesini08}
for a detailed investigation of random and systematic uncertainties in computing the stellar mass function
at $1.3<z<4.0$ (e.g.~the impact of different SED-modeling assumptions,
cosmic variance, and photometric redshift errors).

\subsection{Assigning Gas}
In order to reasonably assign gas to the central galaxies within our halos, we
quantify observationally-inferred relations between the ratio of cold gas mass to stellar mass ($M_{\rm gas}/M_{\rm star}$)
 as a function of stellar mass using the empirical results of
 \citealp{McGaugh05} (blue triangles, for disk-dominated galaxies at $z=0$)
and \citealp{Erb06} (red triangles, for UV-selected galaxies at $z\sim2$), as shown in
the middle and right panels of Figure \ref{McGaughErbfits}.  Though both of these samples
are biased with respect to blue (gas-rich) galaxies, we argue below that by adopting these
relations we are not strongly biasing our overall results.

As shown by the black solid lines in the right panels of Figure \ref{McGaughErbfits},  the $z\sim0$ and $z \sim 2$ cold gas fraction
data can be characterized by a relatively simple function of stellar mass
and redshift:
\begin{equation}
\frac{\Mgas}{\Mstar} = 0.04 \left( \frac{\Mstar}{4.5 \times 10^{11} M_{\odot}} \right)^{- \alpha(z)},
\label{eq:gasmass}
\end{equation}
where the gas fraction relation evolves and steepens with redshift as $\alpha(z)=0.59 (1+z)^{0.45}$.
Assuming a Gaussian scatter about the best-fit lines (independent of mass),
we find that the scatter evolves with redshift as
$\log_{10}{[\sigma(z)]} = 0.34 - 0.19 \log_{10}{(1+z)}$, such that
the correlation between the cold gas fraction and stellar mass is tighter at $z\sim2$ than it is at $z=0$.
The black long-dashed and dotted lines in the right panels of Figure 1 demonstrate
the  $1 \sigma$ and $2 \sigma$ scatter, respectively.
In order to assign gas content to our halos as a function of stellar mass,
we draw randomly from a Gaussian distribution with the average value and
standard deviation given by the above analytic characterizations of $\Mstar(\Mgas)$.

For comparison, the middle panel of Figure \ref{McGaughErbfits} also shows several
additional observational estimates for the gas-fraction relation as a function of stellar mass.
The gold squares and gold X's present the average relation measured by
\cite{Kannappan04} for blue galaxies and red galaxies, respectively.
The  green short-dashed line shows the average (statistical) relation derived
using a combination of published galaxy stellar mass functions and the observed stellar mass-metallicity relation
by \cite{Baldry08}, who used robust chemical evolution arguments to derive implied gas fractions
as a function of stellar mass.  Finally, the average results from direct measurements by
\cite{Wei09}
are shown by the  magenta dot-dot-dot-dashed line.
For the sake of comparison, we show the {\em predicted} relation from the
semi-analytic model of \citealp{Somerville08} (red dot-dashed line).
While we do not utilize these additional
data sets in \emph{constructing} our fitting function, they conform well to our average relation and
certainly lie within the $1 \sigma$
scatter of our fit to the \cite{McGaugh05}  data
(with the exception of low-mass red galaxies from \citealp{Kannappan04}; see discussion below).
We also note that the cold gas fractions derived by \cite{Wright08} for six galaxies at
 $z\sim1.6$ are also consistent
with the evolution in our fit, with every galaxy in their sample
falling within our $2\sigma$ scatter.  Their sample does
have a slightly higher average gas fractions at fixed $M_{\rm star}$ than our adopted relation, but
the discrepancy is not significant given the small-number statistics.
We have also compared our fit to the $34$ galaxies at $z\simeq0.6$ studied in \cite{Hammer09b}, in which
the authors use K-band magnitudes to estimate total stellar mass via the methodology of \cite{Bell03}, and
assume the Schmidt-Kennicutt law to derive gas masses from star formation rates.  Encouragingly,
$26$ of their galaxies ($\sim75\%$) fall within the $1\sigma$ contours of our best-fit at this redshift, with
all $34$ of them within $2\sigma$.

The fact that we have fit our $z=0$ relation to disk-dominated galaxies
 introduces a potential worry about applying the relation
to every galaxy halo in our simulation, including ones that presumably host massive (spheroidal) galaxies.
 However, it is unlikely that this bias will drastically affect our results,
primarily because the gas fractions in the adopted relation are only appreciable
($ \gtrsim 0.5$), in the smallest galaxies (at $z=0$)
with $M_{\rm star} \lesssim 10^{10.5} \Msun$ -- the stellar mass regime that is known to be dominated
by disk-dominated galaxies (see, e.g., the left panel of Figure 2).
For larger galaxies, it is reassuring to note that the average relation for the
red galaxy sample from \cite{Kannappan04} lies within our adopted $\sigma$ scatter and
is in relatively good agreement with the other (disk-selected) observations.  It is
only for \emph{less} massive galaxies (a regime where blue disk galaxies dominate the
total population anyway) where the red galaxy sample of \cite{Kannappan04} becomes significantly
discrepant from our fiducial relation.
Finally, even if our fiducial relation is biased to be slightly high for massive galaxies, the gas fractions
are already small enough that we would never classify them as ``gas-rich'' in our discussions below.

A similar point of concern may be applied to the \cite{Erb06} data at $z \sim 2$.  These galaxies
were selected based on UV luminosity and thus constitute an actively star-forming population.
However, there is a good deal of evidence that UV luminosity is tightly correlated
with total stellar mass (or halo mass) at $z\gtrsim2$
\cite[see e.g., discussion in][and references therein]{Conroy08}.
For example, galaxies with higher UV luminosities
at $z\sim2$ are more strongly clustered \citep{Adelberger05}, suggesting that they reside
within more massive dark matter halos.  In addition,
the UV and V-band luminosity functions of galaxies at $z\sim3$ are in relative agreement,
producing similar number densities for $\sim L^{\ast}$ galaxies
(\citealp{Shapley01,Sawicki06}; also see Table 2, and discussion, in \citealp{Stewart09a}).
As such, it is reasonable to consider a galaxy sample selected on UV luminosity to
contain a fairly representative sample of bright galaxies at $z\sim2$.

We also note that the gas estimates we adopt from \cite{Erb06} assume
the global Schmidt law of
\cite{Kennicutt98}: $\Sigma_{\rm SFR} \propto \Sigma_{\rm gas}^{1.4}$.
Both observations and recent hydrodynamic simulations have suggested that while
this relation is tightly correlated for \emph{molecular} gas, it may underestimate
the \emph{total} gas content, especially for galaxies where the fraction of gas in
molecular form is not uniform \citep[e.g.][]{WongBlitz02,RobertsonKravtsov08,Gnedin08}.
As a consequence, the gas fraction estimates from \cite{Erb06} may represent a lower limit,
such that the evolution of gas fraction with redshift may actually be \emph{steeper}
than our adopted relation.  Insofar as issues of gas accretion
and disk survival are concerned, our relation may be considered a conservative lower limit
on the estimated gas content of our galaxies.

Finally, there has been some discussion in the literature that the stellar initial mass function (IMF) may
evolve systematically to become more top heavy  at high redshift in
galaxies with extremely low metallicities
\citep[e.g.][]{Lucatello05,Tumlinson07,vanDokkum08,Komiya08}.
This evolution in the IMF has not been corrected for in our adopted mapping between
halo mass and stellar mass.  While we do not expect this to significantly affect our
results, including this evolution of the IMF would decrease our stellar mass
estimates at fixed halo mass---which would, in turn, \emph{increase} our estimated
gas fractions.  As such, so far as issues of gas accretion and disk survival are concerned, the
results we present are a conservative lower limit.

With the above qualifications in mind, we now turn to the implications of this
empirically-motivated stellar mass and gas mass assignment prescription.

\begin{figure*}[t!]
  \includegraphics[width=0.32\textwidth]{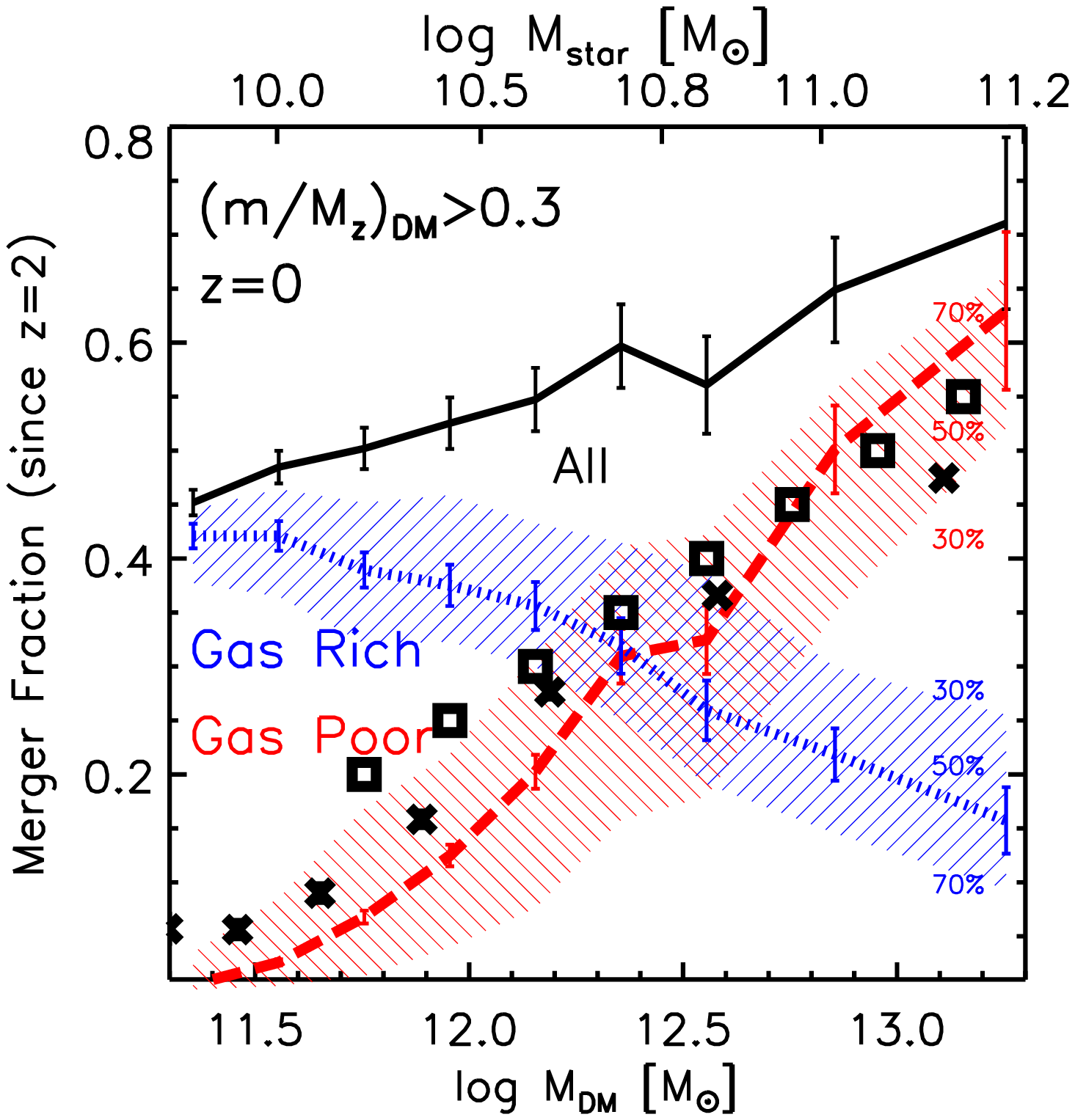}
  \includegraphics[width=0.32\textwidth]{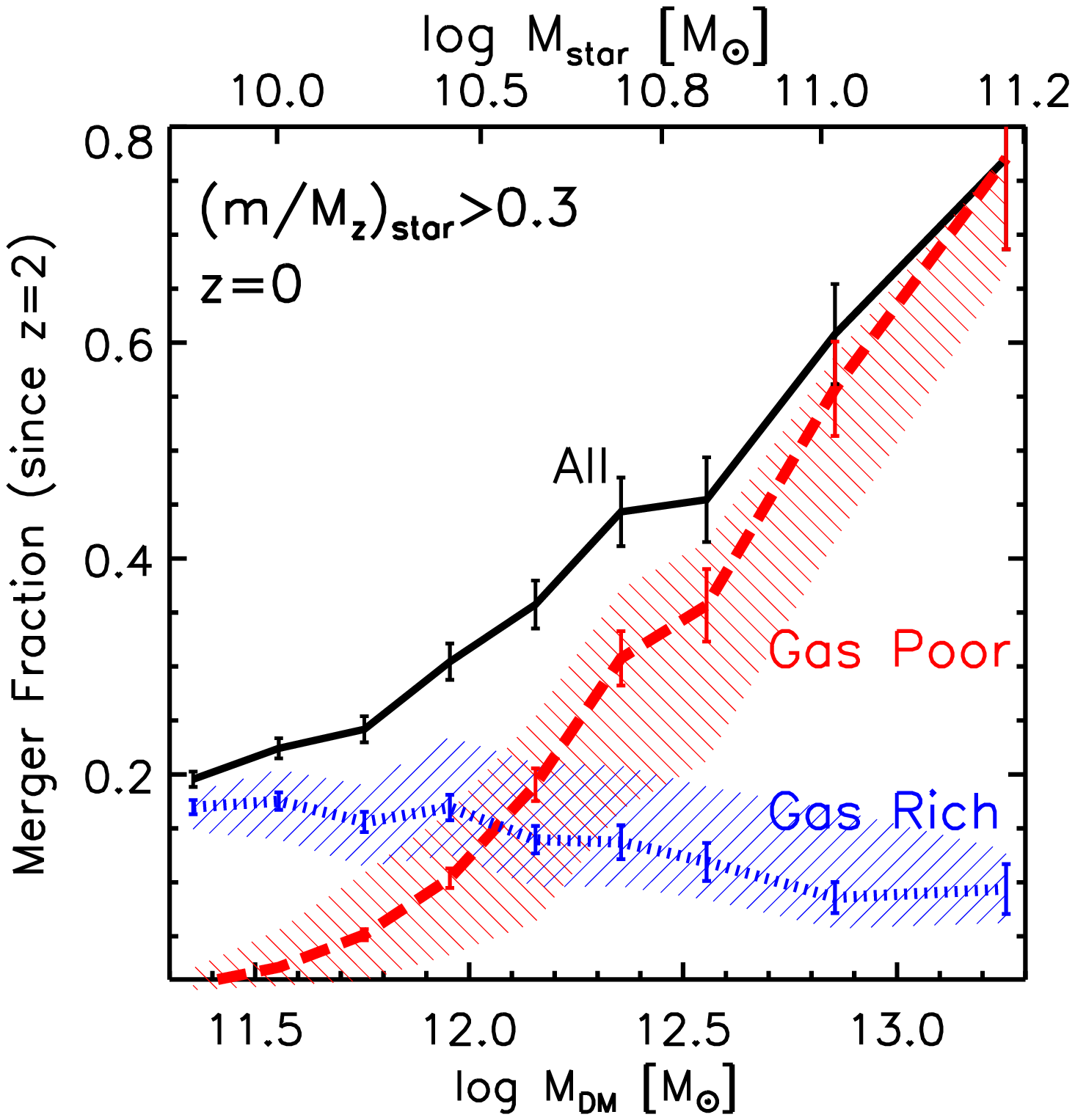}
  \includegraphics[width=0.32\textwidth]{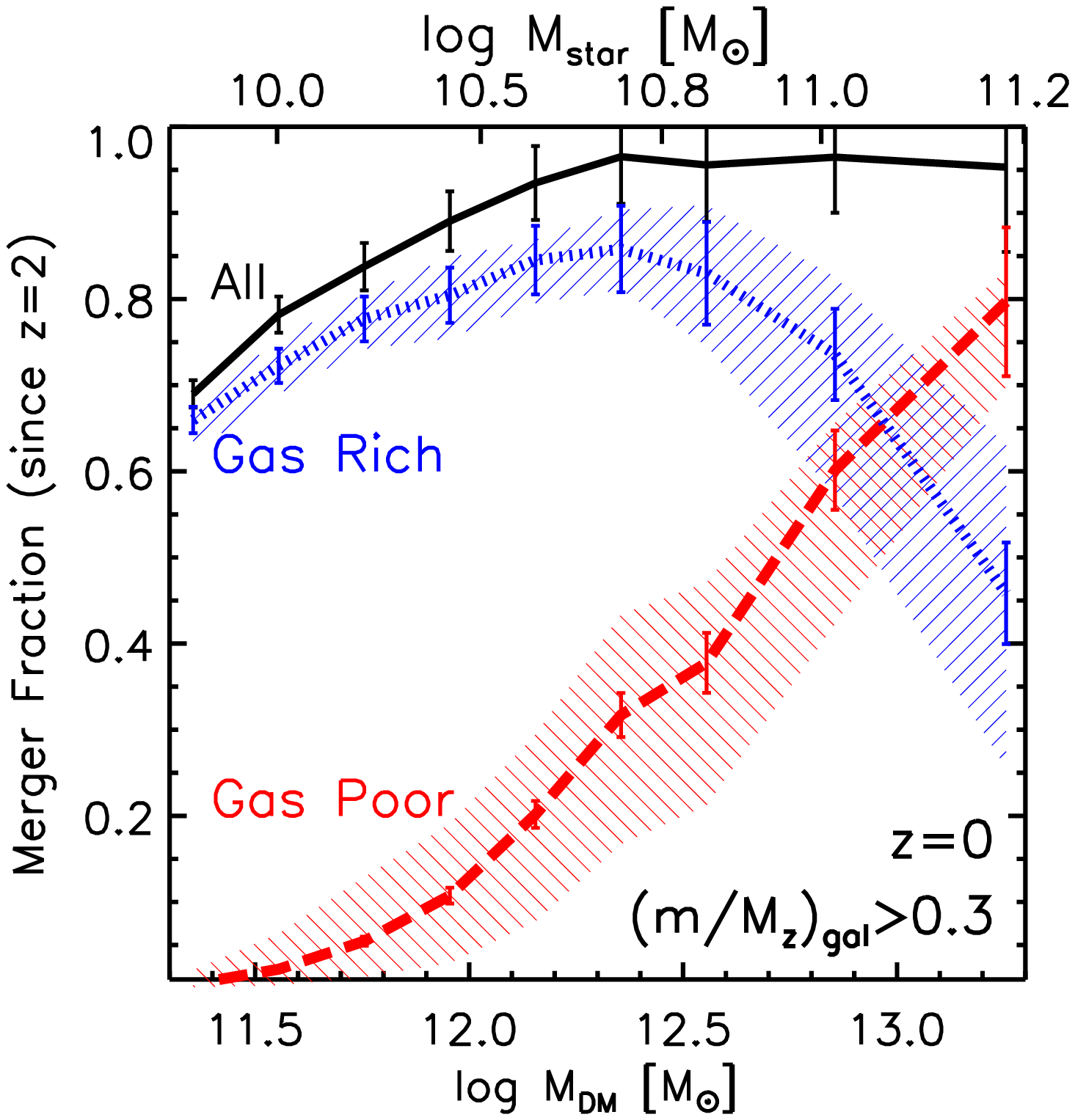}
    \caption{Fraction of dark matter halos with at
least one major merger since $z=2$, as a function
of host halo mass $\Mvir$ (lower axis) and stellar mass $\Mstar$(upper axis),
for varying definitions of `major merger.'
\emph{Left:} Major merger defined by the ratio of dark matter halo masses, $\rDM$.
The black squares and crosses in this figure show the observed early-type fraction as a function of
halo mass from \cite{Weinmann06,Weinmann09}.
\emph{Middle:} Major merger defined by the ratio of the stellar masses in each central galaxy, $\rstar$.
\emph{Right:} Major merger defined by ratio of the total baryonic mass of the central galaxies, $\rgal$.
In each panel, the solid (black) line shows the total merger fraction.
The dashed (red) line shows the merger fraction while \emph{only} considering
mergers between two halos that \emph{both} contain gas-poor central galaxies ($f_g<50\%$).
The dotted (blue) line shows only mergers for which both galaxies are gas-rich ($f_g>50\%$).
The shaded regions surrounding the red and blue lines represent the impact of
varying our distinction between gas-rich and gas-poor from $30\%$ to $70\%$.
The bottom axis shows the same range in halo mass in each panel, while the corresponding stellar (or
baryonic) mass of the central galaxy is shown on the top axis.  Note that the range in y-values in the right panel
is larger than the left and middle panels.
Error bars are Poissonian based on the number of host halos and the total number of mergers,
and do not include possible errors in assigning stars and gas to halos
(though we do account for \emph{scatter} in the $\Mstar(\Mgas)$ relation, see \S2.2,2.3).}
\label{fracz2}
\end{figure*}

\section{Results and Implications}
\subsection{Galaxy Morphology}

We start by investigating the
merger histories of $z=0$ dark matter halos.
The solid black line in the left panel of Figure \ref{fracz2} shows the fraction of dark matter halos that have
experienced at least one major dark matter merger with $\rDM > 0.3$ since $z=2$ as a function of
dark matter halo mass (lower axis label).  Equivalently, the merger fraction as a function of galaxy stellar mass
can be seen by focusing on the upper axis label.
Compare this result to the black squares, which show the early-type fraction
for SDSS galaxies as a function of central halo mass as derived by
Weinmann et al. (2006; with ``early-type'' based on galaxy color and
specific star formation rate) \nocite{Weinmann06} \footnote{Note that in W06,
they divide galaxies into three categories instead of two; early-type, late-type and
intermediate-type.  In order to compare our simple bimodal model to their findings, we
count half of their intermediate-types as early-type, and half as late-type.}.
Also compare to the early-type fraction for SDSS galaxies as a function of halo mass, where
``early-type'' is defined by the concentration parameter ($C>3$), shown as the black crosses,
from \cite{Weinmann09}  (We will refer to these two results as W06 and W09, respectively).
Clearly the fraction of halos with major mergers greatly
exceeds the early-type fraction at low masses.

Consider now the likely baryonic makeup of these mergers.
The (blue) dotted line shows the fraction of halos that have experienced a
\emph{gas-rich} major merger since $z=2$ and the (red) dashed line shows
the fraction of halos with at least one \emph{gas-poor} merger.
In our fiducial case, we define a merger to be gas-rich if
both the central galaxy \emph{and} the infalling satellite galaxy
have more baryonic mass in the form of gas than in stars: $f_g \equiv \Mgas/(\Mgas+\Mstar) > 50\%$.
Similarly, gas-poor mergers are defined such that each of the
progenitors has $f_g < 50 \%$.
The shading of the red and blue bands correspond to varying the definition
of gas-rich from $f_g > 30 \%$ to $f_g > 70 \%$.
\footnote{
Because the morphological impact of a mixed merger (where one galaxy is gas-rich
and one is gas-poor) is largely unclear, we choose to focus on the extreme cases where both are
either gas-rich  or gas poor, and to leave
a more detailed exploration of mixed mergers for future work.
This means that the combined gas-rich fractions and gas-poor fractions in
Figure \ref{fracz2} need \emph{not} equal the total merger fraction, which includes all major
mergers regardless of baryonic content.  However, because a galaxy's gas fraction is a strong function of
halo mass at fixed redshift (and because we define a strict cutoff between gas-rich and gas-poor
based on gas fraction) we find that mixed mergers are less frequent than mergers between two gas-poor or
two gas-rich systems.  If the larger galaxy
in a major merger is gas-rich (by our definition), then the smaller galaxy is
most likely gas-rich as well.  Conversely,
if the smaller galaxy in a major merger is gas-poor, then the larger galaxy is most likely also gas-poor.
While there does exist a characteristic mass scale for which
mixed mergers become a significant portion of the overall merger fraction,
even at this special mass scale they still only constitute about
half of all mergers (see Figure \ref{fractau}, discussion in \S3.3).}

Remarkably, if one makes the simplistic assumption that only gas-poor mergers
generate early-type galaxies (red dashed line) and gas-rich mergers
preserve disks, then the observed SDSS relation from W06 and W09 is reproduced
fairly well.
Specifically, we find that the fraction of halos with a disk-destructive merger
increases from only $\sim 15\%$ at $10^{12} \Msun$ to $\sim 55 \%$ at $10^{13} \Msun$,
 in remarkably good agreement with W06 and W09.  Not only does this
agreement provide a possible solution to disk survivability
within Milky Way-sized halos, but it also implies
that the gas-poor merger history of a dark matter halo may be closely tied to the
halo mass--morphology relation (across this range in halo mass).
Although halo merger rates have been shown to depend strongly on environment
\citep{FakhouriMa09}, suggesting a possible connection to the morphology--density relation
of galaxies,
we see in Figure \ref{fracz2} that the overall halo major merger rate (solid black line)
is \emph{not} steep enough to account for the observed change
in morphological fraction with halo mass.

It is also worth mentioning that
the implied transition between mostly gas-rich and mostly gas-poor mergers occurs
at a characteristic mass $M_{\rm star} \simeq 5 \times 10^{10} \Msun$
(or equivalently $ M_{\rm DM} \simeq 2 \times 10^{12} \Msun$), which is close to the
characteristic bimodality scale that separates
blue, star-forming, disk-dominated systems and
red, bulge-dominated systems with old populations
\citep[typically $\Mstar \sim3 \times 10^{10}\Msun$, see e.g.][]{Kauffmann03,Baldry04,Kannappan04,Baldry06,Cattaneo06,DekelBirnboim06}.

Of course, for detailed treatments of galaxy morphology, this model is too simplistic.  The inclusion of gas-richness
in the efficacy of major mergers to disrupt morphology seems to greatly relieve the
problem of disk stability, and the agreement between the
observed early-type fraction from SDSS and the fraction of halos with at least one gas-poor major merger
is quite remarkable.  However, this is only a first step in understanding the distribution of
galaxy morphologies.  In detail, nothing we have investigated here can explain the prominence of
``bulgeless'' galaxies, as simulated gas-rich major mergers lead to galaxies with noticeable bulges
and disks that are thicker and hotter than the Milky Way.  Even cosmological simulations that produce
thin disk galaxies \cite[e.g.][]{Governato08} require significant smooth gas accretion from the hot halo
after the most recent gas-rich merger in order to form a thin disk.  Such intricate details are
beyond the scope of this paper, as we are primarily concerned with providing
the most robust predictions possible, only using $N$-body dark matter halo merger trees and empirical
relations between $\Mvir, \Mstar,$ and $\Mgas$ to determine gas-rich and gas-poor merger statistics.

Any model that predicts detailed bulge-to-disk mass ratios or
estimates the thinness or thickness of the galactic disk resulting from a merger event must require further assumptions
about the detailed morphological effects of any given merger event, which are still relatively uncertain.
We refer the reader to \cite{Hopkins09} for a more detailed galaxy formation model that
generates bulge and disk mass estimates due to merger events as a function of merger mass ratio, gas fraction
and orbital parameters.  Using a semi-empirical assignment of gas and stars to dark matter halos
similar (but distinct) from our own treatment, they find that cosmologically motivated merger trees
lead to consistent distributions of B/T (bulge mass to total galaxy mass) values as a function of
halo mass and redshift.

\subsection{Alternative definitions for major merger}

The middle and right panels of Figure 2 explore how the implied merger fraction trends
change when one chooses to define major mergers using the stellar-mass ratio, $\rstar>0.3$, and
 total baryonic galaxy mass ratio, $\rgal>0.3$, respectively, rather than the total mass ratio in dark matter.
 Clearly, the implied trends between merger fraction and galaxy halo mass depend sensitively on
whether dark matter mass ratios, stellar mass ratios, or baryonic galaxy mass ratios are considered \citep[also see][]{Maller08, Stewart09conf}.
As seen by the solid black line in the middle panel, high stellar-mass ratio events are rare in small galaxy halos and common in high mass halos.
This follows directly from the fact that low-mass halos tend to have a higher stellar-mass to dark matter mass ratios (see Figure 1).
The trend changes dramatically when the full baryonic mass of the galaxy is considered in the ratio (right panel).  In this case, even small galaxy
halos are expected to have had common mergers with galaxies of a comparable total baryonic mass (note that the range of the
vertical axis has changed in the right-hand panel).  It is clear from this
comparison alone that most of the major mergers experienced by small galaxies must be gas-rich.

We note that the fraction of systems that have experienced at least one gas-rich merger
(and consequently, the total merger fractions as well) show qualitatively different behavior depending on these
definitions.  Gas-rich \emph{halo} mergers ($\rDM>0.3$) are relatively frequent for $M_{\rm DM}=10^{11.5}\Msun$
systems ($40\%$ since $z=2$), with a smoothly declining merger fraction for increasing halo mass (roughly linear in $\log{\Mvir}$), while
the gas-rich \emph{stellar} merger fractions decline in a qualitatively similar fashion but are
universally less common, with merger fractions $<20\%$ for $M_{\rm DM}=10^{11.5}\Msun$.  In contrast, the fraction of halos with
at least one major \emph{galaxy} merger ($\rgal>0.3$) shows completely different behavior, with extremely high
fractions ($40-90\%$) and non-monotonic evolution with $\log{\Mvir}$ (with a maximum value at $M_{\rm DM}\sim10^{12.2}\Msun$).

The behavior of \emph{gas-poor} merger fractions, on the other hand, remains remarkably similar in each case.
Regardless of these three merger ratio definitions, the fraction of halos which have experienced a gas-poor
major merger is negligible at small halos masses ($M_{\rm DM}\sim10^{11.5}\Msun$) and increases roughly linearly
with $\log{\Mvir}$ to a fraction of $65-75\%$ at $M_{\rm DM}=10^{13.2}\Msun$.  Because these gas-poor merger fractions
appear somewhat independent of the merger ratio definition used (and they remain consistent with observed
morphological fractions as a function of halo mass) we again suggest that a dark matter halo's
gas-poor merger history may be a particularly useful tracer of galaxy morphology---more so than the
merger history of all mergers.

One might be tempted to conclude that the total major merger rate in the middle panel (major stellar mergers)
is sufficiently steep to account for the change in morphological fraction with halo mass reported by W06 and W09,
without bothering to account for the gas content of these mergers.
Encouragingly, only $\sim30\%$ of $10^{12}\Msun$ halos have experienced a $\rstar>0.3$ merger since $z=2$.
It is important to note  that even relatively minor, $\rDM=0.1$ dark matter halo mergers
with {\em negligible} stellar content $\rstar \sim 0.03$ are capable of heating and thickening a galactic disk beyond the
properties of the Milky Way \citep{Purcell08b}.  (However, a galaxy need not contain a disk as thin as the Milky Way in order to be
classified as late-type in either W06 or W09, so systems similar to those studied in \cite{Purcell08b} would still be
labeled as ``surviving'' disks in our simple model.)  Still, it is important to keep in mind that
major dark matter mergers do not necessarily correspond to
major stellar mergers, especially at halos less massive than the Milky Way \citep[see e.g.][]{Maller08,Stewart09conf}.
For example, consider a $\Mstar=10^{10} \Msun$ galaxy experiencing
a stellar merger at $z=0, 1, 2$ that is just below our definition of ``major,'' with $\rstar=0.2$.  The corresponding dark matter halo ratios
of these merger events will be $\sim 0.4, 0.4,$ and $0.5$, respectively.  Even if we consider a more minor
stellar merger, with $\rstar=0.1$ at these redshifts, such merger events still correspond to
major dark matter mergers, with dark matter halo ratios of $\sim 0.3, 0.3, 0.4$, respectively.
If gas content is ignored, a substantial fraction of these events will
easily be capable of destroying disk morphologies altogether (not simply thickening and heating the existing
disk).  Still, these mergers
would be classified as ``minor'' stellar mergers, with a stellar merger ratio $<0.3$, even while the total
dark matter mass ratios may approach $2:1$.
In addition, the opposite effect occurs at large galaxy masses.
For example, a major stellar mergers into a $\Mstar=10^{11}\Msun$ galaxy with a $\rstar=0.3$ at $z=0$ only
corresponds to a dark matter halo mass ratio of $\sim0.1$, which is less likely to be morphologically destructive.
This is why the total merger fraction in the middle panel (at high galaxy mass) exceeds that in the left panel:
major stellar mergers only correspond to minor dark halo mergers at this mass regime.
Thus, we conclude that the major stellar merger fractions in the middle panel likely present an uneven, and potentially
incomplete picture of possible means of disk destruction via mergers.

\subsection{Gas Delivery Via Mergers}

\begin{figure*}[t!]
  \includegraphics[width=0.49\textwidth]{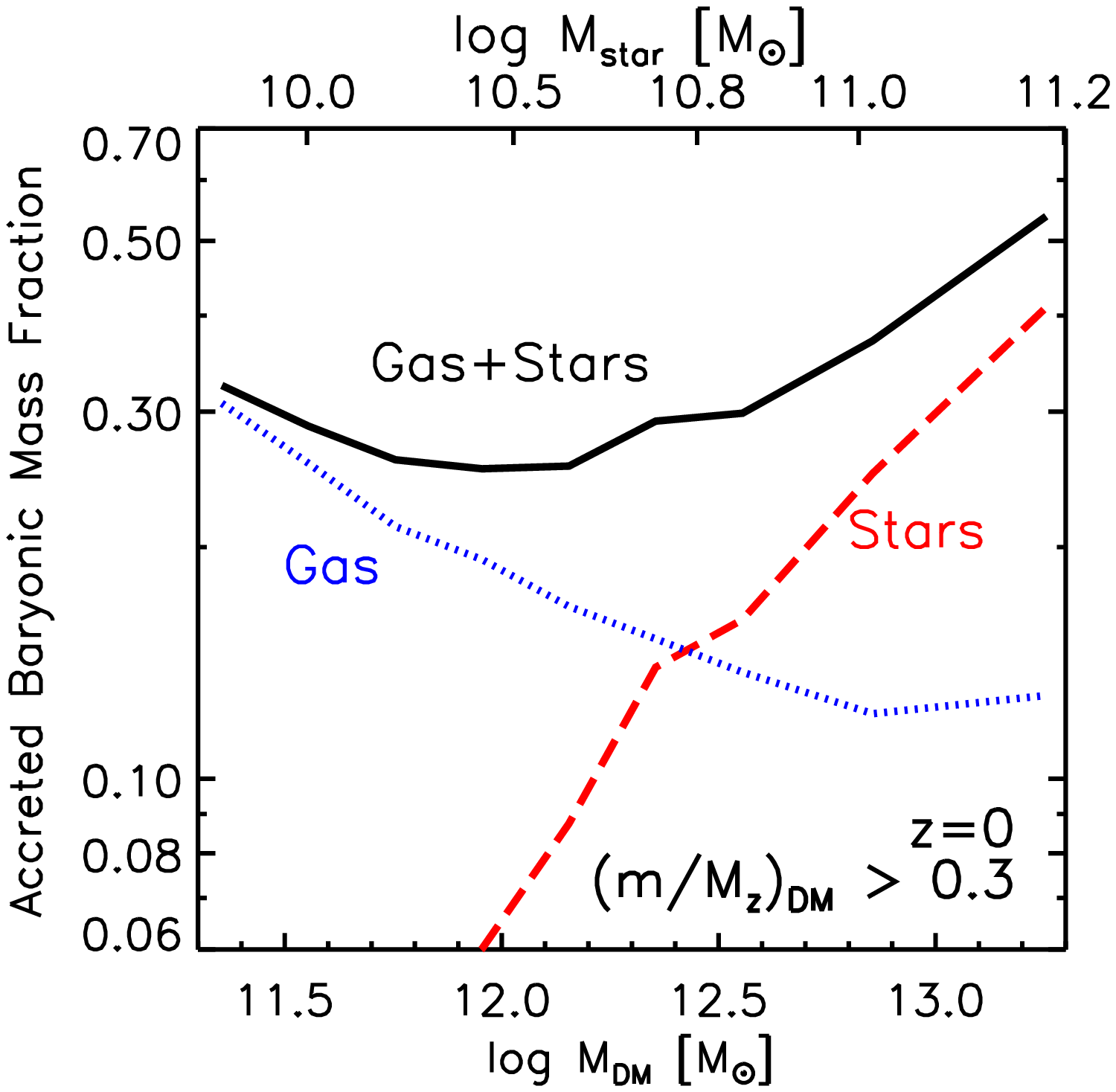}
  \includegraphics[width=0.49\textwidth]{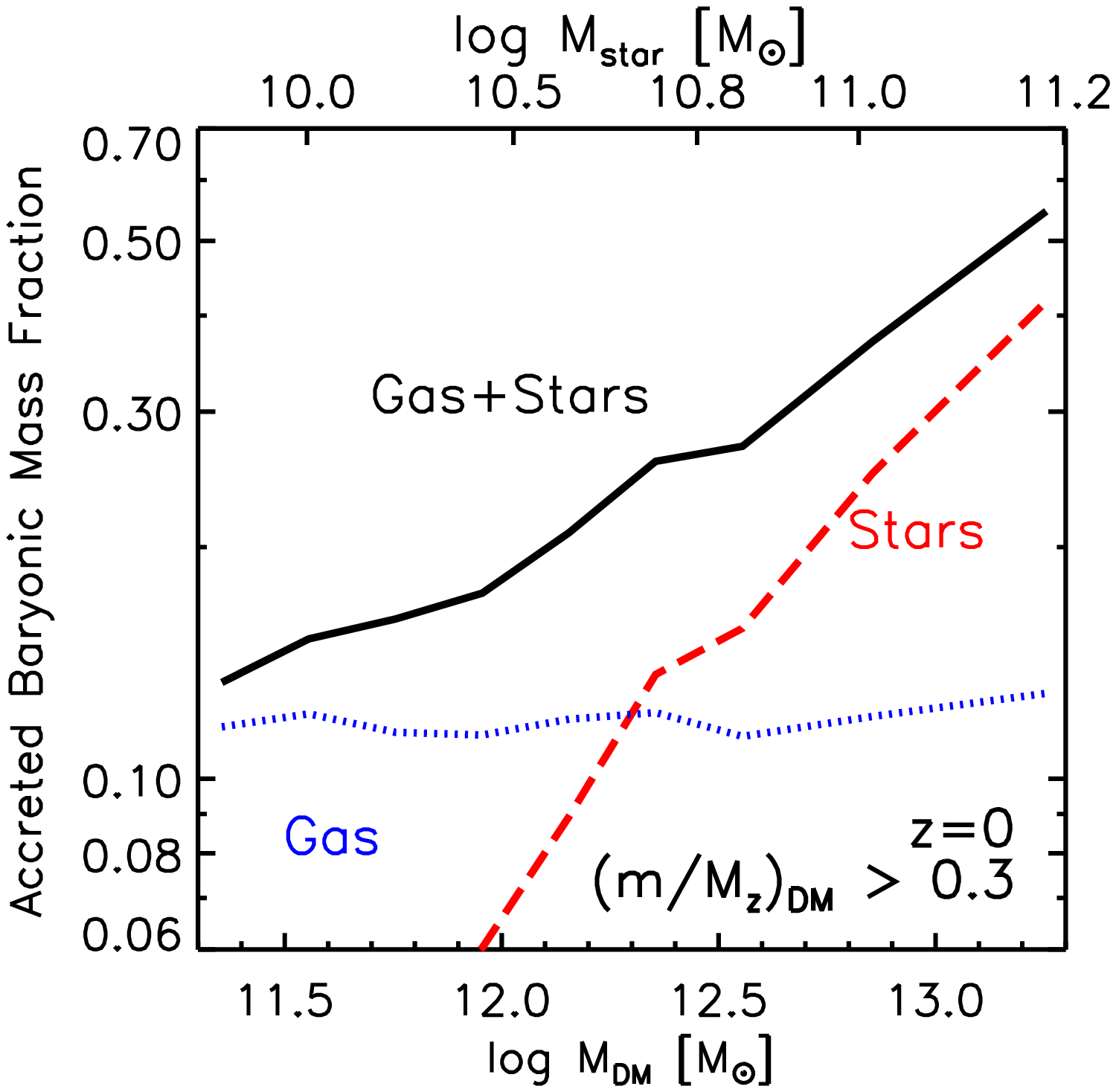}
    \caption{The fraction of a $z=0$ galaxy's total baryonic mass that was accreted directly via
major mergers or moderately sized mergers since $z=2$, as a function of halo mass.
\emph{Left:} model A, in which we assign gas by Equation \ref{eq:gasmass}, but impose an upper limit
due to the universal baryon fraction, such that $\Mgas+\Mstar\leq f_b \Mvir$.
\emph{Right:} model B, in which we assign gas by Equation \ref{eq:gasmass}, but impose an upper limit
such that $\Mgas\leq f_{lim}\Mvir$, where $f_{lim}$ at each redshift is the value of $\Mgas/\Mvir$ when
$\Mstar=3\times10^8 \Msun$ at that redshift.
In both panels, the dotted (blue) and dashed (red) lines show the
accreted baryonic mass fraction from gas and stars, respectively, while the solid (black) line
shows the total.}
\label{baryonacc}
\end{figure*}

In \cite{Stewart09a}, we discuss the observational implications of two well-known consequences of galaxy halo mergers:
merger-induced starbursts and morphological disturbance.
A third potentially important consequence of mergers is
the direct, cumulative deposition of cold baryons (gas and stars) onto galaxies.
We are now concerned with the exact baryonic content of each merging galaxy, whereas we have previously only focused
on whether or not a given halo lies above or below an arbitrary gas fraction threshold.  As such, we impose an additional
constraint to our method for assigning gas (as outlined in \S2).  Specifically, when estimating the gas content of halos at high
redshift, blindly extrapolating equation \ref{eq:gasmass} to arbitrarily small stellar masses
sometimes results in more baryons in a given halo than the universal baryon fraction of matter in the Universe.
This is an unphysical situation that arises from extrapolation far beyond the regime where $\Mgas(\Mstar)$ is well-constrained.
In order to avoid unphysically high gas content of low mass halos at high redshift, we present two models for assigning
upper limits to the gas content of low stellar mass galaxies.  In the model A (left panel of figure \ref{baryonacc}),
we set an upper limit on equation \ref{eq:gasmass} such that $\Mgas+\Mstar \leq f_b \Mvir$, where $f_b=0.17$.
In model B, we define the upper limit by the ratio of gas mass to halo mass: ($f_{lim} \equiv \Mgas/\Mvir$
at $\Mstar=3\times10^8\Msun$).  For galaxies with stellar mass lower than this threshold, we then set
$\Mgas = f_{lim} \Mvir$.  This model, which always assigns less (or equal) gas per galaxy than the first model,
is used to construct the right panel of figure \ref{baryonacc}.

In both panels of Figure \ref{baryonacc}, the solid black lines show the fraction of a central galaxy's
current ($z=0$) baryonic mass ($M_{\rm gal} = \Mgas+\Mstar$) acquired directly via major mergers
as a function of halo mass $f_{\rm merged} \equiv M_{\rm merged}/M_{\rm gal}(z=0)$.  Specifically, $M_{\rm merge}$ includes
all of the baryonic mass in mergers obeying $\rDM > 0.3$ since $z=2$.
Focusing on the left panel (model A: $\Mgas+\Mstar \leq f_b \Mvir$)
the first clear result is that the merged baryonic fraction is significant:
$\sim 30-50\%$  of the final galaxy mass is accreted directly
in the form of major mergers. Given the effectiveness of dynamical friction in major mergers,
we expect the majority of the accreted baryonic material in these events to be deposited in the central galaxy itself.  In principle,
this limits the fraction of a galaxy's baryons that can be acquired by direct hot halo cooling, cold flows, or minor mergers
to  $50-70\%$, depending on the halo mass of interest. Of course, (gaseous) baryons deposited via major mergers could in principle be blown out by energetic feedback, but all in all,
this result on major merger deposition (much like known results on cold flows)
would seem to make the 'overcooling' problem in galaxy formation more difficult.
While the fraction of baryons accreted directly as cold gas via major mergers is less substantial
in the right panel (model B: $\Mgas \leq f_{lim}\Mvir$), we still find significant accretion:
$15-50\%$ of the final galaxy mass.

The dotted (blue) and dashed (red) lines in Figure \ref{baryonacc}
separate the total baryonic accretion fraction from major mergers into
contributions from gas and stars, respectively (this is not a division between gas-rich and gas-poor mergers as before, but
rather an integrated accounting of all material regardless of the makeup of the merged progenitors).
We see that in both models, the baryonic accretion onto smaller halos ($\Mvir\lesssim10^{12.3}\Msun$) is
typically dominated by the gas content of the infalling galaxies, while the baryonic makeup of merged material
into more massive systems is dominated by
the infalling galaxies' \emph{stellar} content.
For Milky Way-size systems ($\Mvir=10^{12}\Msun$), we find in model A (B) that typically $\sim30\% (20\%)$ of
a galaxy's baryonic content was accreted in the form of gas and stars directly via major mergers, with most
of this accretion dominated by cold gas.
In both models, more massive systems ($\Mvir=10^{13}\Msun$) typically accrete $\sim30\% (10\%)$ of their baryons
as stars (gas) via major mergers, with no noticeable discrepancies between the two models.
Though not shown, we find that most baryonic accretion from major mergers ($70-80\%$ of stars, $50-70\%$ of gas) occurred
at later times.

How do our results change if we include more minor mergers in our accounting?
If we count up all of the baryonic acquisition in mergers larger than
$\rDM>0.1$, we find that the mass fraction accreted as stars is boosted by a factor of
of $\sim1.5$ from the panels shown, while the accreted gas is amplified by a factor of $\sim1.7$, both roughly
independent of halo mass.
We caution, however, that the importance of minor mergers in delivering baryons to central galaxies is significantly less
clear than it is with major mergers.
While the baryons associated with major mergers almost certainly become deposited
directly onto the central galaxy (see e.g. the simulations of Purcell et al. 2008b)
the ultimate fate of the baryons in minor mergers will depend sensitively
on the orbital properties of the secondary and on the potential presence of hot gas halo
around the primary galaxy.    Past work has demonstrated that
the {\em  stellar} material in minor mergers will likely contribute
to extended diffuse light components like stellar halos or intracluster light \citep[e.g.,][]{bj05, Purcell07, CWK07, Purcell08},
but the destiny of accreted {\em gas}  (which is the dominant component for galaxy halos)
in these minor mergers is relatively unexplored.
One possibility is that the gas in minor mergers is quickly liberated via ram pressure stripping
\citep[see, e.g.][]{Grcevich08}
and that it either evaporates into the hot halo itself or eventually rains down onto the
galaxy, possibly in the form of high-velocity clouds.  These interesting possibilities are clearly beyond
the scope of the present work but provide important avenues for future investigation.

The relation between gas mass and stellar mass below $\Mstar=3\times10^8\Msun$ remains relatively uncertain,
making it impossible to know which panel of this figure is a more accurate representation of galaxy formation.
In detail, we expect our two models to bracket reasonable expectations for the true relation between $\Mstar$ and
$\Mgas$ in this regime.  Nevertheless, focusing on model A for the time being (left panel), we find it
interesting that there appears to be a minimum in merged baryon fraction at the $\Mvir \sim 10^{12} \Msun$ scale, which
corresponds closely to the well-known mass scale of maximum galaxy formation efficiency ($\sim L_*$
in the galaxy luminosity function).  Although it is unclear whether this minimum exists in reality, or is merely
an artifice of the manner in which we have assigned gas, we speculate on a possible correspondence between
galaxy formation efficiency and the merged baryon fraction.  Specifically, a minimum in the merged baryon fraction naturally implies a
maximum in the baryon fraction accreted via smooth gas accretion from the hot halo or from cold streams.
We speculate that smooth gas accretion might allow for a higher efficiency in star formation than accreting large
clumps of gas via major mergers, because gas-rich mergers are likely to trigger massive starbursts
that may blow significant gas content out of the central galaxy.  We also speculate that accretion of
stars via mergers may also be inefficient, since some fraction of a satellite galaxy's stars is likely to be
distributed into the stellar halo before reaching the central galaxy within massive halos.
If this is the case, that smooth gas accretion forms stars most efficiently, then it would be reasonable
to expect a maximum in the baryon fraction of smoothly accreted gas to correlate with the
maximum galaxy formation efficiency.  However, we emphasize that this possible correlation between
minimum merged baryonic fraction and maximum star formation efficiency is not a robust prediction of our model,
but merely a speculation (for example, model B results in no such minimum in the merged baryonic fraction).

\begin{figure*}[tb!]
  \includegraphics[width=0.98\textwidth]{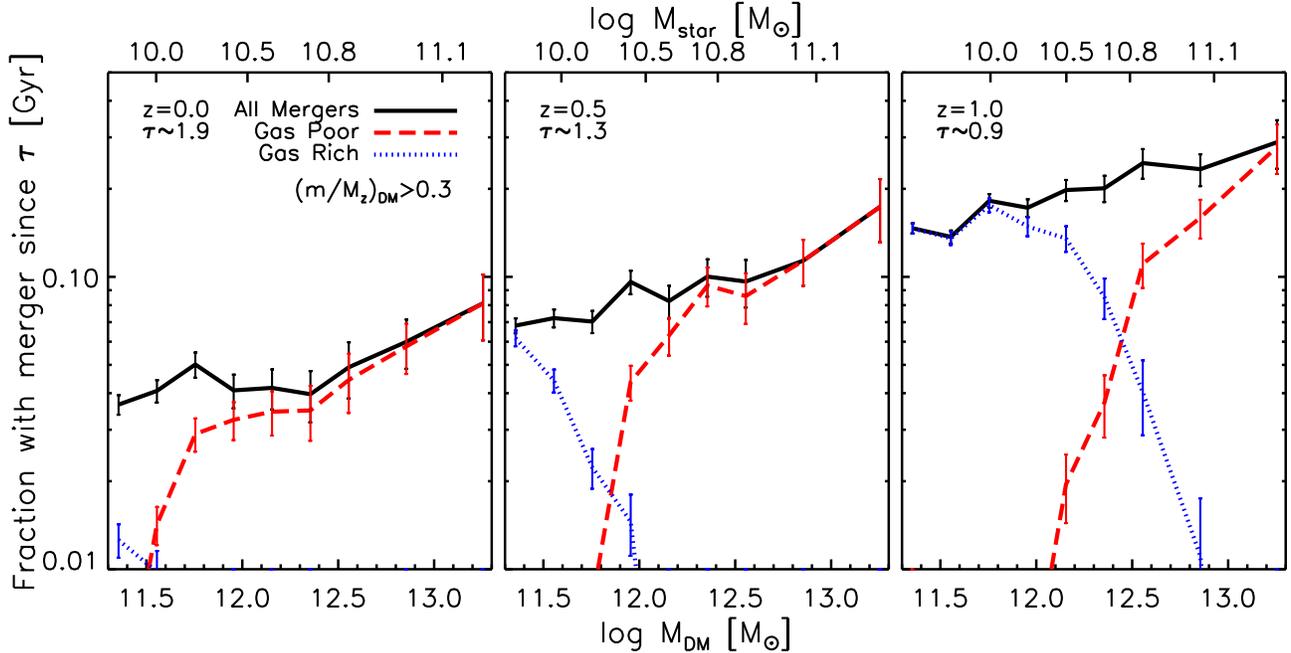}
   \caption{Major merger fraction within the past dynamical time of the halo, $\tau$,
 as a function of halo mass. The solid (black) line shows the total merger fraction for
dark matter halos (no baryons included).  The dashed (red) line shows the
merger fraction while only considering gas poor mergers ($f_g<50\%$).
The dotted (blue) line shows only gas-rich mergers ($f_g>50\%$).
The three panels show results for $z=0$, $z=0.5$, and
$z=1$, for which the halo dynamical time, $\tau\simeq1.9, 1.3, 0.9$ Gyr, respectively.
We primarily focus on the decomposition of these merger fractions into gas-rich and
gas-poor mergers.  We refer the reader to \cite{Stewart09a} for a
detailed analysis of the dependence of dark matter
of merger rates and merger fractions on redshift, halo mass, and mass ratio.}
\label{fractau}
\end{figure*}

\subsection{Redshift Evolution}

While the cumulative fraction of halos that have \emph{ever}
experienced a gas-rich or gas-poor major merger (since $z=2$) is the most pertinent
question for morphological evolution and disk survivability (see \S3.1), another
point of interest is the redshift evolution of a more instantaneous measure of the merger
rate of gas-poor and gas-rich mergers.  Figure \ref{fractau} shows the fraction
of halos that have experienced at least one major merger with $\rDM>0.3$ in the past halo dynamical time,
$\tau$~\footnote{As in \cite{Stewart09a}, in which we studied the evolution
of the halo merger rate with redshift, we again adopt
$\tau(z) = R/V \propto (\Delta_v(z) \, \rho_u(z))^{-1/2}$,
such that the halo dynamical time evolves with redshift, but is independent of halo mass.}.
As in Figure \ref{fracz2}, the solid (black) line shows the total merger fraction for dark
matter halos, while the dotted (blue) and dashed (red) lines show the major merger fraction for
gas-poor ($f_g<50\%$) and gas-rich ($f_g>50\%$) mergers, respectively.
Because the quantitative values of these merger fractions depend sensitively on the merger
timescale in question at each redshift, we choose to focus on two
important \emph{qualitative} results from this
figure.  We refer the reader to \cite{Stewart09a} for a more detailed discussion of the evolution
of the halo merger rate (and merger fractions) with redshift, halo mass, and merger mass ratio
\citep[see also][for the effects of halo environment on the merger rate]{FakhouriMa09}.

The first feature of note in this figure
is the presence of a typical transition mass, $(M_t)_{\rm DM}$, such that
most of the recent major mergers into halos less massive than $(M_t)_{\rm DM}$
are gas-rich, while most of the recent major mergers into halos more massive than $(M_t)_{\rm DM}$ are gas-poor.
The existence of this transition mass is primarily due to
the strong dependence of galaxy gas fractions on stellar mass (and thus, halo mass)
at fixed redshift.  In addition to creating this transition mass, the dependence of
gas fraction on halo mass also
results in a very limited mass range for which mixed mergers
(where one galaxy is gas-poor and the
other is gas-rich) constitute a significant portion of all mergers (at most $\sim50\%$).
This effect is apparent in Figure \ref{fractau} where the combined total of the
gas-poor and gas-rich merger fractions fall significantly short of the total.
Not unexpectedly, this range of importance for mixed mergers is centered on $(M_t)_{\rm DM}$.

The other important result from Figure \ref{fractau}
is that $(M_t)_{\rm DM}$ is more massive at higher redshifts,
with $(M_t)_{\rm DM}\simeq 10^{11.4}, 10^{11.9}, 10^{12.8} \Msun$ at $z=0.0, 0.5, 1.0$, respectively.
This arises naturally from the strong increase in galaxy gas fractions to higher redshift.
The corresponding galaxy stellar mass transitions at $z=0.0, 0.5,$ and  $1.0$,
are $(M_t)_{\rm star} \sim 10^{9.7},10^{10.3},10^{11.0} M_{\odot}$ (upper horizontal axis in Figure 4).
 Of course, the precise value of $M_t$ at each redshift will depend to some degree
on our definitions of ``gas-rich'' and ``gas-poor,'' but the \emph{existence} of this
transition mass, and its qualitative evolution with redshift should be robust to
changes in these definitions.

Consider  recent major mergers into Milky-Way size
$10^{12}\Msun$ halos.  At $z=0$, mergers of this kind are very uncommon.  Only
 $\sim 5 \%$ of Milky-Way size halos should have experienced a major dark matter accretion event
 with $\rDM >0.3$ in the last
 $\tau \sim 2$ Gyr.  However,  when these major mergers do occur at $z=0$ they are
very likely gas-poor ($\sim 0.04/0.05 = 80 \%$ of the time).
 On the other hand major mergers are fairly  common in $10^{12} \Msun$ halos at
$z \sim 1$, with  $\sim 15 \%$ experiencing such a merger in the last $\tau \sim 1$ Gyr.
Nevertheless,  these higher redshift mergers are almost universally gas-rich.
Under the presumption that
gas-rich mergers do not destroy disk morphologies, the evolution of the merger rate with redshift and
in the associated gas-rich transition mass makes it increasingly likely that major mergers {\em build} disk galaxies at high
redshift rather than destroy them (c.f. Robertson et al. 2006a, Robertson \& Bullock 2008).
If we were to boldly extrapolate our trends to higher redshift,
we would expect that nearly \emph{all} major mergers into
halos with $\Mvir<10^{12}\Msun$ should be gas-rich ($f_g>50\%$) at $z>1$.

 Encouragingly, \cite{Lin08} observe a similar redshift evolution between
gas-rich and gas-poor mergers
by studying the close-pair counts of galaxies from the DEEP2
Redshift Survey.  While our definitions vary in detail from theirs (they divide galaxy pairs
into wet and dry mergers based on galaxy colors, and bin their sample by total galaxy luminosity, while
we use galaxy gas fractions to define gas-poor versus gas-rich mergers, and bin by galaxy stellar mass)
they also find that at fixed luminosity (stellar mass)
the percentage of major mergers that are dry (gas-poor) should decrease with increasing redshift,
while the percentage of major mergers that are wet (gas-rich) should increase.
We reserve a more detailed comparison to their results for a future study,
but we find the qualitative agreement encouraging.

\subsection{Comparison to Previous Work}

Recent studies of galaxy formation at high redshift using hydrodynamic simulations have
stressed the importance of smooth accretion of cold gas from filamentary streams.
For example,  \cite{Keres08} compared the accretion rate of gas
onto galaxies via cold flows and via mergers, and found that at $z=1-2$, only about half of
all gas accretion (onto galaxies corresponding to $\Mvir\gtrsim10^{11.3}\Msun$)
is in the form of mergers, (where gas from mergers was defined as any gas
that was added to galaxies in dense baryonic clumps).  Similarly,
\cite{Dekel08} found that half of cold gas infall onto massive $z=2$ galaxies is acquired
via mergers with $\rDM>0.1$, and the other acquired from cold flows.  Indeed, even
studies that focus primarily on the importance of galaxy mergers also note that
smooth gas accretion is at least as dominant as galaxy mergers in the mass
buildup of galaxies \citep[e.g.][]{maller06}.  Our results do not contradict these
expectations.  As demonstrated in
Figure \ref{baryonacc}, we expect that $\sim 30-40\%$ of a typical galaxy's baryons
should have been accreted directly via major mergers with $\rDM > 0.3$.  This leaves significant room for
cold-flow gas to contribute to the baryonic assembly of small galaxies and for cooling to contribute
to the buildup of larger galaxies.
Though, as mentioned above, we do expect that the percentage of merger-delivered baryons
could rise to as much as $\sim 60\%$ if all of the baryonic material from $\rDM>0.1$
mergers is able to find its way into the central galaxy.

\cite{Brooks08} used a high-resolution cosmological hydrodynamic simulation to study
the gas accretion onto four disk galaxies within halos of masses $10^{10.7-12.7} M_{\odot}$,
and also found that smooth accretion of gas (either shocked or un-shocked) dominates the mass buildup of
their galaxies.  When comparing smooth gas accretion to gas infall from
mergers (using a generous definition of what qualifies as a merger) they found that
$\sim25\%$ of the total gas infall into their Milky Way-size galaxy derives from
mergers, with
$\sim10\%$ of the final stellar content at $z=0$ being accreted directly as stars from
mergers.  While their detailed results (and definitions) differ slightly
from our own, the rough consistency between their simulation and our own semi-empirical
approach is quite encouraging.

One point of caution associated with the discussion of cold flows is that these
predictions are based entirely on simulations that do not generally reproduce the observed
baryonic mass function and stellar mass function of galaxies.  It is possible that the
cold flows are somehow restricted in the real universe in a way that solves the well-known
over-cooling problem in galaxy formation.   Due to the inherent difficulties in
detecting cold filaments of gas locally and at high redshift, there has yet to be an observational
confirmation of a star forming galaxy fueled by the smooth accretion of cold gas along
filaments, as seen in hydrodynamic
simulations.  In contrast, our predictions for the accretion of stars and gas via major mergers is solidly normalized
against observations, and is arguably inevitable in the context of LCDM merger histories.
Of course, baryonic material (especially gas) that is delivered via major mergers
need not remain in the central galaxy indefinitely.  Gas accreted either along cold flows or through
major mergers may be subsequently expelled via supernovae or AGN feedback
\citep[e.g.][]{Benson03,DiMatteo05,Springel05b,Somerville08}.
In this respect, Figure \ref{fractau} represents an upper limit on the baryonic contribution from
major mergers, with respect to the total amount of barons \emph{currently} exist in the galaxy.

\section{Conclusion}
\label{Conclusion}

We have used dark matter halo merger trees from a large cosmology $N$-body simulation
together with observationally-normalized relationships between
dark matter halo mass, galaxy stellar mass, and galaxy gas mass to explore the baryonic
content of galaxy mergers back to redshift $z=2$.  Though our adopted associations between halo
mass and the baryonic content of galaxies cannot be precisely correct, it is almost
certainly accurate in its scalings with halo mass and redshift, and has the added advantage that
it is \emph{independent of any uncertain galaxy formation physics}.  Indeed, any self-consistent galaxy formation
model that is set within the LCDM framework would certainly need to reproduce our gross baryonic assignments
in order to reproduce the observed universe.  Our main results based on this methodology
may be summarized as follows:

\begin{enumerate}

\item The vast majority ($\sim 85\%$) of the major mergers experienced by Milky-Way size galaxies since $z=2$ should
    have been gas-rich, and this fraction drops significantly towards higher mass systems (see Figure 2).
    Remarkably, the fraction of galaxies with gas-poor major mergers matches well to the observed fraction of bulge-dominated
    galaxies as a function of halo mass from $\Mvir=10^{11}$ to $10^{13}\Msun$.

\item Though {\em recent} major mergers are expected to be rare for small galaxies in the local universe, the recent mergers
    that do occur should typically be gas poor.  At higher redshift, recent mergers become more common and the probability that
    such a merger is gas-rich also increases (see Figure 4).  One can define a transition dark matter halo mass
    $M_t$, below which most of the recent major mergers are
    gas-rich and above which they are gas poor, and  this transition
    mass increases with redshift: $(M_t)_{\rm DM} \sim 10^{11.4}, 10^{11.9}, 10^{12.8}\Msun$ at
    $z=0.0,0.5,1.0$.  As a result, the vast majority of recent major mergers into galaxy-size
    $\Mvir < 10^{12}\Msun$  dark matter halos are expected to be gas-rich at $z < 1$.

\item A significant fraction ($20-50\%$) of the baryonic mass in field galaxies at $z=0$ should have been deposited directly
      via major mergers since $z=2$.  For less massive galaxies, $\Mvir \sim 10^{11.5}\Msun$, the vast majority
      of the merger-acquired baryons are gaseous, while in more massive galaxies
      $\Mvir\sim10^{13}\Msun$, major mergers bring in mostly stars (see Figure 3).  For Milky Way-size
      systems, major mergers since $z=2$ bring in $\sim 30\%$ of the galaxy's $z=0$ baryonic mass, with most of this
      contribution in the form of gas.

\end{enumerate}

Many of these conclusions lend support to the conjecture of Robertson et al. (2006a) and \cite{Brook07b}, who were the
first to forcefully suggested a scenario where
gas-rich mergers play an important role in building and stabilizing disk galaxies at high redshift.  Though our conclusions are far from
a sufficient test of this idea, we have demonstrated that gas-rich mergers should be common enough to make it viable for serious consideration.

Among our most interesting results is the similarity between our predicted gas-poor merger fraction with halo mass
and the observed early-type galaxy fraction with halo mass (Figure 2, left panel).
Of course, even if gas-rich mergers do preserve disks, there are
many openings for concern.  For example, the current presentation leaves
 little room for the production of
bulge-dominated systems by means \emph{other} than major mergers.
In an extreme yet illustrative example, \cite{Bournaud07} used a suite
of focused simulations to show that bulge-dominated galaxies may be formed by
successive minor mergers.  Disk galaxies can also grow massive bulges by
secular processes (typically bulges formed in this way show kinematically distinct
properties from classical bulges, and are referred to as ``pseudobulges'')
\citep[e.g.][]{Courteau96,KormendyKennicutt04,Kormendy05,Kormendy08}.

An interesting possibility in this context of disk survival and secular evolution
is that we have been too conservative in our
classification of `gas-rich'.  Our fiducial division between gas-rich and gas poor at $f_g = 50 \%$ was motivated by the idealized
simulations studied by Robertson et al. (2006a) and Hopkins et al. (2008).  However
\cite{Governato08} used a cosmologically self-consistent hydrodynamic simulation
to demonstrate the creation of spiral galaxy at $z=0$ within a system
that experienced a very major ($\rDM>0.8$) merger at $z=0.8$.
The two progenitor galaxies in this case were only moderately gas-rich ($f_g \sim 20\%$).
Despite these relatively low gas fractions, the merger remnant was able to quickly reform a disk
via the cooling of gas from the hot phase.  If we use this result as
motivation to focus on the more lenient ($f_g>30\%$) definition of gas-rich in
Figure \ref{fracz2}, our gas-poor merger fractions drop to $10-20\%$ smaller than
the observed bulge-dominated fractions, leaving room for processes other than
gas-poor major mergers to cause a significant portion of morphological transformations.

The general semi-empirical findings we have presented here
may be regarded as accurate (not precise) predictions based on
merger histories of LCDM halos and observed relations.
As such, it is reassuring that our almost unavoidable qualitative trends
are consistent with a growing body of work that stresses the importance of gas-richness in preserving
disk morphologies during mergers
\citep{Barnes02,Brook04,SpringelHernquist05,Robertson06a,Brook07a,Brook07b,
Governato07,Governato08, Hopkins08g,RobertsonBullock08}.
Although we have focused primarily on issues of morphological transformation,
disk survival, and baryonic accretion via mergers in this paper,
we believe that in future work, the semi-empirical approach we have used here
may provide a useful tool in exploring a vast array
of galaxy properties and evolutionary mechanisms.

\acknowledgements The simulation used in this paper was run on the
Columbia machine at NASA Ames.  We would like to thank Anatoly Klypin
for running the simulation and making it available to us.  We are also
indebted to Brandon Allgood for providing the merger trees.
We thank Charlie Conroy for sharing his abundance matching data
so we could assign stellar masses to dark matter halos, and Lisa
Wei (and collaborators) for sharing gas fraction data from
an upcoming paper.  We also than the anonymous referee, whose suggestions
helped improve the quality of this paper.
JSB and KRS are supported by NSF grant AST 05-07916.  RHW
was supported in part by the U.S. Department of Energy under contract
number DE-AC02-76SF00515 and by a Terman Fellowship from Stanford
University.  AHM
acknowledges partial support from a CUNY GRTI-ROUND 9 grant and from
an ROA supplement to NSF grant AST 05-07916.
JSB and KRS are supported by the Center for Cosmology at the
University of California, Irvine.

\bibliography{ms}

\end{document}